\documentclass[12pt]{article}

\usepackage{amsmath}
\usepackage{graphicx}
\usepackage{enumerate}
\usepackage[round]{natbib}
\usepackage{url} 

\usepackage{bm}
\usepackage{amsfonts,amssymb}
\usepackage{mathrsfs}

\newtheorem{Definition}{Definition}

\begin{document}

  \title{\bf The Likelihood Ratio Test and Full Bayesian Significance Test under small sample sizes for contingency tables}
  \author{Natalia L. Oliveira\thanks{\textit{This work was partially supported by the Brazilian agencies FAPESP grant 2012/16669-4, and CNPq grants 302046/2009-7 and 308776/2014-3. The funders had no role in study design, data collection and analysis, decision to publish, or preparation of the manuscript.}}\hspace{.2cm}\\
    Department of Statistics, Federal University of Sao Carlos\\
    and \\
    Carlos A. de B. Pereira\\
    Department of Statistics, University of Sao Paulo\\
    and \\
    Marcio A. Diniz\\
    Department of Statistics, Federal University of Sao Carlos\\
    and \\
    Adriano Polpo \\
    Department of Statistics, Federal University of Sao Carlos}
  \maketitle

\begin{abstract}
Hypothesis testing in contingency tables is usually based on asymptotic results, thereby restricting its proper use to large samples. To study these tests in small samples, we consider the likelihood ratio test and define an accurate index, the P-value, for the celebrated hypotheses of homogeneity, independence, and Hardy-Weinberg equilibrium. The aim is to understand the use of the asymptotic results of the frequentist Likelihood Ratio Test and the Bayesian FBST -- Full Bayesian Significance Test -- under small-sample scenarios. The proposed exact P-value is used as a benchmark to understand the other indices. We perform analysis in different scenarios, considering different sample sizes and different table dimensions. The exact Fisher test for $2 \times 2$ tables that drastically reduces the sample space is also discussed. The main message of this paper is that all indices have very similar behavior, so the tests based on asymptotic results are very good to be used in any circumstance, even with small sample sizes. 
\end{abstract}

\noindent%
{\it Keywords:}  Categorical data; e-value; FBST; hypothesis test; p-value; significance test.
\vfill

\section{Introduction}

We discuss indices for homogeneity, independence, and Hardy-Weinberg equilibrium hypotheses \citep{Emigh1980,Delgado2001} in contingency tables. We propose the P-value -- an exact evaluation of the Likelihood Ratio Test (LRT) -- as a benchmark significance index. Based on the work of \citet{Pereira1993}, the idea is to evaluate the probability distribution of all possible tables on the sample space under the hypothesis. Once the distribution of sampling a contingency table under the hypothesis is known, we are able to compute the distribution of the Likelihood Ratio Test (LRT) statistics. The main difficulty is that it is a time-consuming computational procedure, being only feasible for small sample sizes and/or for tables of small dimension. 

The presented P-value of the LRT, a way to calculate an exact inference, is called P-value with capital letter P in order to differentiate it from the asymptotic p-value. The aim is to compare the behavior of the frequentist LRT asymptotic p-value, the LRT exact P-value, the Fisher test exact p-value, the Chi-Square test asymptotic p-value, and the Bayesian asymptotic e-value \citep{Pereira1999,Pereira2008a} and the approximation (Markov Chain Monte Carlo) of the exact e-value. We are interested in the values of the indices, not in the acceptance or rejection of the hypothesis. That is, our focus is on the significance test, which consists of the evaluation of the p-(e-)values. In an applied setting, the researcher can, based on the indices, make his/her decision about his/her problem. We are not interested in comparing the values of the indices with some fixed significance value (generally 5\%) to decide the if the hypothesis should be accepted or rejected. With this goal in mind, all significance indices considered here, including the P-value and the Bayesian e-value, are in agreement with the ASA's statement on significance indices \citep{Wasserstein2016}.

From a historical perspective, hypothesis testing has been the most widely used statistical tool in many fields of science \citep{Lawson2000,Herrmann2007,Montgomery2010}. For categorical data, \cite{Agresti2001} discusses some exact procedures to perform inference. \cite{Agresti2002} presents methodological procedures for hypothesis testing for contingency tables. \citet{Eberhardt1977} compares, under an asymptotic perspective, two tests for equality of two proportions considering Goodman's $Y^2$ and $\chi^2$ statistics. To test the independence of two classifiers in contingency tables, \citet{Pagano1981} presents an algorithm for finding the exact permutation significance level for $r \times c$ contingency tables. \citet{Irony2000b}, studies a simple way to compare two correlated proportions. More recently, \citet{Zhang2012} presents the exact likelihood ratio test for equality of two normal populations.

The Likelihood Ratio Test (LRT) asymptotic p-value \citep{Casella2001}, the Chi-Square test asymptotic p-value \citep{Agresti2007}, Fisher's homogeneity exact test \citep{Agresti2007,Irony1986}, and the Full Bayesian Significance Test (FBST) asymptotic and exact e-value \citep{Pereira1999,Pereira2008a} are presented in detail for the case of $2 \times 2$ contingency tables considering homogeneity hypothesis (Section \ref{sec_homo22}). The homogeneity and independence hypotheses for tables of any dimension and Hardy-Weinberg equilibrium hypothesis are discussed in sections \ref{sec_homo}, \ref{sec_ind} and \ref{sec_hw}.

We study the relationship between indices in Section \ref{sec_compare}. In a similar study, \citet{Diniz2012a} considers continuous random variables using the e-value and the LRT p-value. It is shown that these indices share an asymptotic relationship. In our case, all indices have similar behavior, including in small sample size scenarios. Moreover, the present results are not based on a simulation study; we compute the indices for all possible tables in the sample space. 

In addition to our focus on the study of significance tests, we also provide, for the frequentist indices, a study of power functions to compare the indices for the homogeneity hypothesis ($2\times 2$ tables) and Hardy-Weinberg equilibrium hypothesis (Section \ref{sec_power}). The Fisher exact test was the least powerful, followed by the Chi-Square test, the exact LRT (P-value) and the asymptotic LRT, the most powerful one. We did not evaluate the power function for the FBST; firstly, because it is not the aim of the Bayesian paradigm, and secondly, to do so, it would be necessary to define a decision rule for the FBST, which is not the scope of this paper. We also note that, under the hull hypothesis, considering the significance level 5\%, all frequentist indices achieved 5\% rejection. Section \ref{sec_final} presents our final comments.

\section{Significance indices}

\subsection{Homogeneity test for $2 \times 2$ contingency tables}
\label{sec_homo22}

\begin{figure}[!ht]
\centering
  \includegraphics[width=0.45\textwidth,keepaspectratio=true]{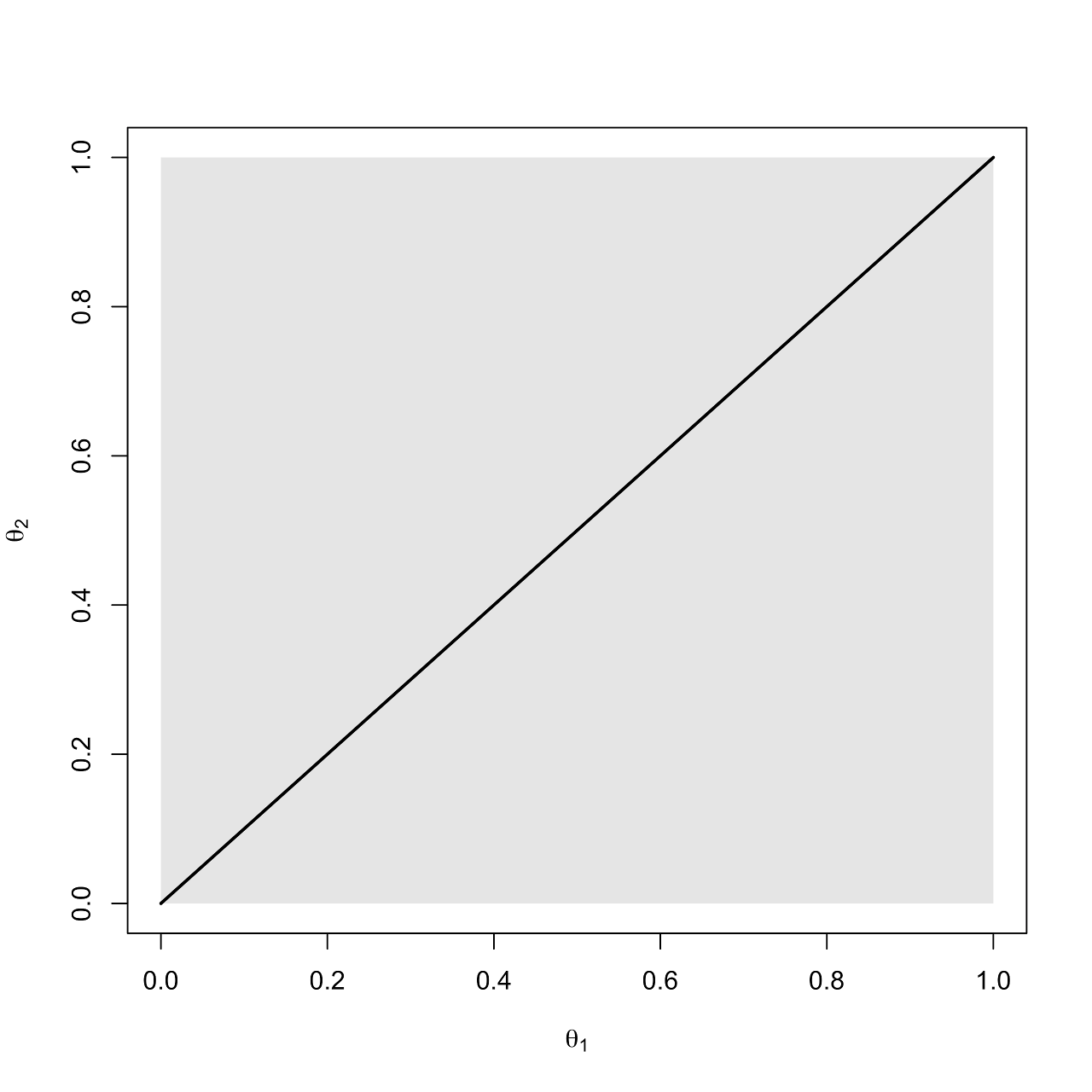}
\caption{Homogeneity hypothesis (black line) for $2 \times 2$ contingency table and parameter space (gray shading). \label{fig_1}}
\end{figure}

Let $X_1$ and $X_2$ be two random variables, represented in Table \ref{tab_cont}, $x_{11}$ and $x_{21}$ being their observed values, and $n_{1 \cdot}$ and $n_{2\cdot}$ fixed sample sizes. Consider the distributions of $X_1$ and $X_2$ as $\textnormal{Binomial}(n_{1\cdot}, \theta_{11})$ and $X_2$ a $\textnormal{Binomial}(n_{2\cdot}, \theta_{21})$ for describing the chances of a subject belong to category $C_1$ in two distinct populations being compared. Both populations are partitioned into two categories $C_1$ and $C_2$ and the object is to test homogeneity among the two unknown population frequencies, $\bm{H}: \theta_{11} = \theta_{21} = \theta$. This hypothesis is geometrically represented in Figure \ref{fig_1}. 

\begin{table}[ht]
\centering
\caption{\label{tab_cont} Contingency table $2 \times 2$.}
\begin{tabular}{cccc}
\hline
  & $C_1$ & $C_2$ & total \\
\hline
$X_1$ &  $x_{11}$  &  $x_{21}$  & $n_{1\cdot}$ \\
$X_2$ &  $x_{21}$  &  $x_{22}$  & $n_{2\cdot}$ \\
\hline
\end{tabular}
\end{table}

The likelihood function is specified by
\begin{equation}
L(\theta_{11},\theta_{21} \mid x_{11},x_{21},n_{1\cdot},n_{2\cdot}) = \frac{n_{1\cdot}! n_{2\cdot}!}{x_{11}! x_{21}! x_{12}! x_{22}!} \theta_{11}^{x_{11}} \theta_{21}^{x_{21}} (1-\theta_{11})^{x_{12}} (1-\theta_{21})^{x_{22}}, 
\end{equation}
where $ 0 \leq \theta_i \leq 1, i=1,2$. Under ${\bm{H}}$, the likelihood function simplifies to
\begin{equation}
\label{eq_vH}
L(\theta \mid x_{11},x_{21},n_{1\cdot},n_{2\cdot},\bm{H}) = \frac{n_{1\cdot}! n_{2\cdot}!}{x_{11}! x_{21}! x_{12}! x_{22}!} \theta^{x_{11}+x_{21}} (1-\theta)^{x_{12}+x_{22}}, 0 \leq \theta \leq 1,
\end{equation}
and the LRT test statistics is:
\begin{equation}
\label{eq_estat}
\lambda({x_{11},x_{21}}) = \frac{\sup\limits_{\theta \in \Theta_{\bm{H}}} L(\theta_{11},\theta_{21} \mid x_{11},x_{21},n_{1\cdot},n_{2\cdot})}{\sup\limits_{\theta \in \Theta} L(\theta_{11},\theta_{21} \mid x_{11},x_{21},n_{1\cdot},n_{2\cdot}) } =  \frac{(\frac{x_{11}+x_{21}}{n_{1\cdot}+n_{2\cdot}})^{x_{11}+x_{21}} (\frac{x_{12}+x_{22}}{n_{1\cdot}+n_{2\cdot}})^{x_{12}+x_{22}}}{ \left(\frac{x_{11}}{n_{1\cdot}} \right)^{x_{11}} \left(\frac{x_{12}}{n_{1\cdot}} \right)^{x_{12}} \left(\frac{x_{21}}{n_{2\cdot}} \right)^{x_{21}} \left(\frac{x_{22}}{n_{2\cdot}} \right)^{x_{22}}},
\end{equation}
in which $\Theta_{\bm{H}}$ is the parametric set defined by the hypothesis. 

~\\
$\bullet$ {\bf{P-value:}}\\
To define the P-value, we use the predictive distributions of $X_1$ and $X_2$ before any data were observed. The proposed P-value is an alternative way to calculate an exact p-value for the LRT. The goal is to find a distribution for the contingency table under $\bm{H}$ that is not a function on $\theta$. We consider $\theta$ a nuisance parameter in the likelihood function in (\ref{eq_vH}) and integrate it over $\theta$ in order to eliminate it. That is, 
\begin{eqnarray}
\label{eq_hxy}
h(x_{11},x_{21}) & = & \int_0^1 L(\theta \mid x_{11},x_{21},n_{1\cdot},n_{2\cdot},{\bm{H}}) \textnormal{d}\theta \nonumber \\
& = & \frac{n_{1\cdot}! n_{2\cdot}!}{x_{11}! x_{21}! x_{12}! x_{22}!} \int_0^1 \theta^{x_{11}+x_{21}} (1-\theta)^{x_{12}+x_{22}} \textnormal{d}\theta \nonumber \\
& = & \binom{n_{1\cdot}}{x_{11}}\binom{n_{2\cdot}}{x_{21}} \frac{(x_{11}+x_{21})! (x_{12}+x_{22})!}{(n_{1\cdot}+n_{2\cdot}+1)!} \nonumber \\
& = & \frac{\binom{n_{1\cdot}}{x_{11}}\binom{n_{2\cdot}}{x_{21}}}{\binom{n_{1\cdot}+n_{2\cdot}}{x_{11}+x_{21}}} \frac{1}{(n_{1\cdot}+n_{2\cdot}+1)}.
\end{eqnarray}

To obtain the probability function $\Pr(X_1=x_{11},X_2=x_{21} \mid \bm{H})$, one needs to find a normalization constant. 
\begin{equation}
\label{eq_dist}
\Pr(X_1=x_{11},X_{2}=x_{21} \mid \bm{H}) = \frac{h(x_{11},x_{21})}{\sum\limits_{i=0}^{n_{1\cdot}}\sum\limits_{j=0}^{n_{2\cdot}} h(i,j)}.
\end{equation}
Note that to calculate (\ref{eq_dist}), we evaluate $h(\cdot,\cdot)$ for all possible tables. In the case of a homogeneity hypothesis for $2 \times 2$ contingency tables, $\sum\limits_{i=0}^{n_{1\cdot}}\sum\limits_{j=0}^{n_{2\cdot}} h(i,j) = 1$. We present the table's probability in terms of this sum to obtain a general formula for all hypotheses and table dimensions considered here, since in other scenarios this sum is different from $1$. For example, the sum of $h$ for all possible $2 \times 2$ tables considering independence hypothesis with $n=2$ is $2304$. The P-value calculation follows directly from the test statistic distribution: 
\[ \textnormal{P-value} = \Pr(\lambda(X_1,X_{2}) \leq \lambda(x_{11},x_{21}) \mid \bm{H}) = 
\mathop{\sum \sum}_{(i,j):~ \lambda(X_1,X_{2}) \leq \lambda(x_{11},x_{21})} \Pr(X_1=i,X_{2}=j \mid \bm{H}), \]
in which $\lambda(x_{11},x_{21})$ is the observed test statistic, as in (\ref{eq_estat}).

~\\
$\bullet$ {\bf{Full Bayesian Significance Test:}}\\
Our Bayesian approach is based on the FBST (Full Bayesian Significance Test).

\begin{Definition}
\label{def_e}
Let $\pi(\theta \mid \bm{x})$ be the posterior density function of $\theta$ given the observed sample and $T(\bm{x})=\{\theta \in \Theta : \pi(\theta \mid \bm{x}) \geq \sup_{\theta \in \Theta_{\bm{H}}}\pi(\theta \mid \bm{x})\}$. The supporting evidence measure for the hypothesis $\theta \in \Theta_{\bm{H}}$ is defined as $Ev(\Theta_{\bm{H}},\bm{x})=1-\Pr(\theta \in T(\bm{x}) \mid \bm{x})$.
\end{Definition}

Consider that, \textit{a priori}, $\theta_{11}$ and $\theta_{21}$ are independent and both follow a Uniform$(0,1)$ distribution. Recall that $X_1$ and $X_{2}$ given $\theta_{11}$ and $\theta_{21}$ are Binomial distributed. Hence, the posterior distributions for $\theta_{11}$ and $\theta_{21}$ are independent $\textnormal{Beta}(x_{11}+1,n_{1\cdot}-x_{11}+1)$ and $\textnormal{Beta}(x_{21}+1,n_{2\cdot}-x_{21}+1)$. Under the hypothesis $\bm{H}$, the posterior distribution is
\[ \pi(\theta \mid x_{11},x_{21},n_{1\cdot},n_{2\cdot},{\bm{H}}) = \frac{\theta^{x_{11}+x_{21}} (1-\theta)^{x_{12}+x_{22}}}{\mathcal{B}(x_{11}+1,x_{12}+1) \mathcal{B}(x_{21}+1,x_{22}+1)} \]
and by maximizing it in $\theta$ we obtain $\sup_{\theta \in (0,1)} \pi(\theta \mid x_{11},x_{21},n_{1\cdot},n_{2\cdot},{\bm{H}})$, where $\mathcal{B}(\cdot,\cdot)$ is the Beta function. Since $x_{11}$, $x_{21}$, $n_{1\cdot}$ and $n_{2\cdot}$ are integers, 
\begin{eqnarray*}
\pi(\theta \mid x_{11},x_{21},n_{1\cdot},n_{2\cdot},{\bm{H}}) & = & \binom{n_{1\cdot}}{x_{11}} \binom{n_{2\cdot}}{x_{21}} (n_{1\cdot}+1)(n_{2\cdot}+1) \theta^{x_{11}+x_{21}} (1-\theta)^{x_{12}+x_{22}}, \\
\sup_{\theta \in (0,1)} \pi(\theta \mid x_{11},x_{21},n_{1\cdot},n_{2\cdot},{\bm{H}}) & = & \\
&~&\hspace{-2cm} = \frac{(n_{1\cdot}+1)! (n_{2\cdot}+1)!}{x_{11}!x_{21}!x_{12}!x_{22}!} \left(\frac{x_{11}+x_{21}}{n_{1\cdot}+n_{2\cdot}}\right)^{x_{11}+x_{21}} \left(\frac{x_{12}+x_{22}}{n_{1\cdot}+n_{2\cdot}}\right)^{x_{12}+x_{22}},
\end{eqnarray*}
the hypothesis' tangent set, $T$, is
\begin{multline*}
T(x_{11},x_{21},n_{1\cdot},n_{2\cdot}) = \Big\{(\theta_{11},\theta_{21}) \in (0,1) \times (0,1): \\ \pi(\theta_{11},\theta_{21} \mid x_{11},x_{21},n_{1\cdot},n_{2\cdot}) \geq \sup_{\theta \in (0,1)} \pi(\theta \mid x_{11},x_{21},n_{1\cdot},n_{2\cdot}, \bm{H})\Big\},
\end{multline*}
and 
\[ \textnormal{e-value} = 1-\Pr[\theta \in T(x_{11},x_{21},n_{1\cdot},n_{2\cdot})]. \]

To calculate the approximate e-value, we use the following algorithm:
\begin{enumerate}
\item A random sample of size $k$ is generated from posterior distribution of $\theta_{11}, \theta_{21}$, obtaining $\{\theta_{x_{11} 1}, \theta_{x_{21} 1}\}, \ldots, \{\theta_{x_{11} k}, \theta_{x_{21} k}\}$.

\item The e-value is calculated by
\[ 1-\frac{1}{k} \sum_{i=1}^k I\left(\pi(\theta_{x_{11} i}, \theta_{x_{21} i} \mid x_{11},x_{21},n_{1\cdot},n_{2\cdot}) \geq \sup_{\theta \in (0,1)} \pi(\theta \mid x_{11},x_{21},n_{1\cdot},n_{2\cdot})\right),\]
in which $I(A)$ is the indicator function of set $A$.
\end{enumerate}

~\\
$\bullet$ {\bf{Other indices:}}\\
For the LRT, the statistic $-2\ln[ \lambda(X_1,X_2)]$ has asymptotically a chi-square distribution with $1$ degree of freedom, which is $dim(\Theta)-dim(\Theta_{\bm{H}})$ \citep{Casella2001}. The FBST uses the same statistic, however the asymptotic distribution is a chi-square with $2$ degrees of freedom \citep{Pereira2008a}, which is $dim(\Theta)$. For the chi-square test and the Fisher's exact test for homogeneity see \citet{Agresti2007}.

For the sake of brevity, next section only presents the results since they are similar to the ones of this section.

\subsection{Homogeneity hypothesis for $\ell \times c$ contingency tables }
\label{sec_homo}

Let $\bm{X_{i}}, i=1, ..., \ell$ be random variables that can are represented in Table \ref{tab_vero2} and $n_{1\cdot}, n_{2\cdot}, \ldots, n_{\ell \cdot}$ known constants. 

\begin{table}[ht]
\centering
\caption{\label{tab_vero2} Contingency table $\ell \times c$.}
\begin{tabular}{c|cccc|c}
\hline
    &  $\bm{C_{1}}$       &  $\bm{C_{2}}$       & $\cdots$ & $\bm{C_{c}}$    & total \\
\hline
   $\bm{X_{1}}$ &  $x_{11}$    &  $x_{12}$    &          & $x_{1c}$ & $n_{1\cdot}$  \\
   $\bm{X_{2}}$ &  $x_{21}$    &  $x_{22}$    &          & $x_{2c}$ & $n_{2\cdot}$  \\
$\vdots$ &              &              & $\ddots$ & $\vdots$     & $\vdots$  \\
$\bm{X_{\ell }}$ & $x_{\ell 1}$ & $x_{\ell 2}$ & $\cdots$ & $x_{\ell c}$ & $n_{\ell \cdot}$  \\
\hline
total    &  $n_{\cdot 1}$  &  $n_{\cdot 2}$  &  $\cdots$ & $n_{\cdot c}$ & $n_{\cdot \cdot}$  \\
\hline
\end{tabular}
\end{table}

Assuming that $\bm{X_{i}}, i=1, ..., \ell$, follows a $\textnormal{Multinomial}(n_{i\cdot}, \theta_{i1}, \ldots, \theta_{ic})$ distribution, we are interested in testing if their distributions are homogeneous with respect to categories $\bm{C_{j}}$, $j=1, ..., c$. That is,
\[ \bm{H}: \left\{\begin{array}{rcl}
\theta_1 & = & \theta_{11} = \theta_{21} = \cdots = \theta_{\ell 1},\\
\theta_2 & = & \theta_{12} = \theta_{22} = \cdots = \theta_{\ell 2},\\
      & \vdots & \\
\theta_{c-1} & = & \theta_{1(c-1)} = \theta_{2(c-1)} = \cdots = \theta_{\ell (c-1)},\\
\end{array} \right. \]
\noindent
in which $\theta_{c} = 1 - \sum_{k=1}^{c-1} \theta_k$, $0 \leq \theta_{j} \leq 1$, $\forall j = 1, \ldots, c$.

Let $\bm{x}$ be all observed values presented in Table (\ref{tab_vero2}) and $\bm{\theta}$ all the parameters. The likelihood function is
\[L(\bm{\theta} \mid \bm{x}) = \left[ \left. \prod\limits_{i=1}^\ell n_i! \middle/ \prod\limits_{i=1}^\ell \prod\limits_{j=1}^c x_{ij}! \right. \right] \prod\limits_{i=1}^\ell \prod\limits_{j=1}^c \theta_{ij}^{x_{ij}}, \]
and under the hypothesis $\bm{H}$,
\[L(\bm{\theta} \mid \bm{x}, \bm{H}) = \left[ \left. \prod\limits_{i=1}^\ell n_i! \middle/ \prod\limits_{i=1}^\ell \prod\limits_{j=1}^c x_{ij}! \right. \right] \prod\limits_{j=1}^c \theta_j^{n_{\cdot j}}, \textnormal{where}~~ n_{\cdot j} = \sum_{i=1}^{\ell} x_{ij}. \]

The LRT $\lambda$ statistic is
\begin{equation}
\label{eq_lg}
\lambda(\bm{x}) = \left. \prod\limits_{j=1}^c \left(\frac{n_{\cdot j}}{n_{\cdot\cdot}} \right)^{n_{\cdot j}} \middle/ \prod\limits_{i=1}^\ell \prod\limits_{j=1}^c \left(\frac{x_{ij}}{n_{i\cdot}} \right)^{x_{ij}} \right. .
\end{equation}

~\\
$\bullet$ {\bf{P-value:}}\\
To obtain the P-value, we need the function $h(\bm{x})$. In this scenario, 
\[ h(\bm{x}) = \left. (n_{\cdot \cdot} +c-1)! \prod\limits_{i=1}^\ell n_i! \middle/ \left[ \left( \prod\limits_{i=1}^\ell \prod\limits_{j=1}^c x_{ij}!\right) \left(\prod\limits_{j=1}^c n_{\cdot j}!\right) \right] \right. , \]
and the P-value's calculation follows as in Subsection \ref{sec_homo22}.

~\\
$\bullet$ {\bf{FBST:}}\\
Assuming ~a ~~$\textnormal{Dirichlet}(1,1,\ldots,1)$ ~prior ~for ~$\{\theta_{i1}, \ldots, \theta_{ic}\}$, and ~since $\bm{X_{i}}$ ~follows ~a \mbox{$\textnormal{Multinomial}(n_i, \theta_{i1}, \ldots, \theta_{ic})$} distribution, ~~then ~the ~~posterior ~~distribution ~is ~a \mbox{$\textnormal{Dirichlet}(x_{i1}+1,x_{i2}+1,\ldots,x_{ic}+1)$, $i=1,\ldots,\ell$}.

In this setting, 
\[ \sup_{\bm{\theta} \in \Theta_{\bm{H}}} \pi(\bm{\theta} \mid \bm{x})
= \frac{x_{11}!\cdots x_{1c}!\cdots x_{\ell1}! \cdots x_{\ell c}!}{(x_{1\cdot}+c -1)! \ldots (x_{\ell\cdot}+c -1)!} \left( \frac{x_{\cdot 1}}{n} \right) ^{x_{\cdot 1}} \cdots \left( \frac{x_{\cdot c}}{n} \right) ^ {x_{\cdot c}}, \]
and we can obtain the e-value from Definition \ref{def_e}. 

~\\
$\bullet$ {\bf{Other indices:}}\\
Both ~~asymptotic ~~LRT ~~p-value ~~and ~~asymptotic ~~e-value ~~are ~~calculated ~~as \mbox{$\Pr[-2\ln(\lambda(\bm{X})) \leq -2\ln(\lambda(\bm{x}))]$}, but while the LRT considers that this statistic follows a $\mathcal{X}^2$ distribution with $(\ell-1)(c-1)$ degrees of freedom, the FBST considers that it follows a $\mathcal{X}^2$ distribution with $(\ell \times c)-1$ degrees of freedom. The Chi-Square homogeneity test is also obtained. 

\subsection{Independence hypothesis for $\ell \times c$ contingency tables }
\label{sec_ind}

Consider that $\theta_{ij}$ is the probability of observing a sample in the cell at row $i$ and column $j$, $\theta_{i \cdot}$ is the probability of observing a sample in row $i$, $\theta_{\cdot j}$ is the probability of observing a sample in column $j$, $0 \leq \theta_{ij} \leq 1$, $0 \leq \theta_{i \cdot} \leq 1$, $0 \leq \theta_{\cdot j} \leq 1$, $i = 1, \ldots, \ell$, $j = 1, \ldots, c$, $\sum_{i=1}^{\ell} \sum_{j=1}^{c} \theta_{ij} = 1$, $\sum_{i=1}^{\ell} \theta_{i \cdot} = 1$, and $\sum_{j=1}^{c} \theta_{\cdot j} = 1$.

\begin{figure}
\centering
  \includegraphics[width=0.45\textwidth,keepaspectratio=true]{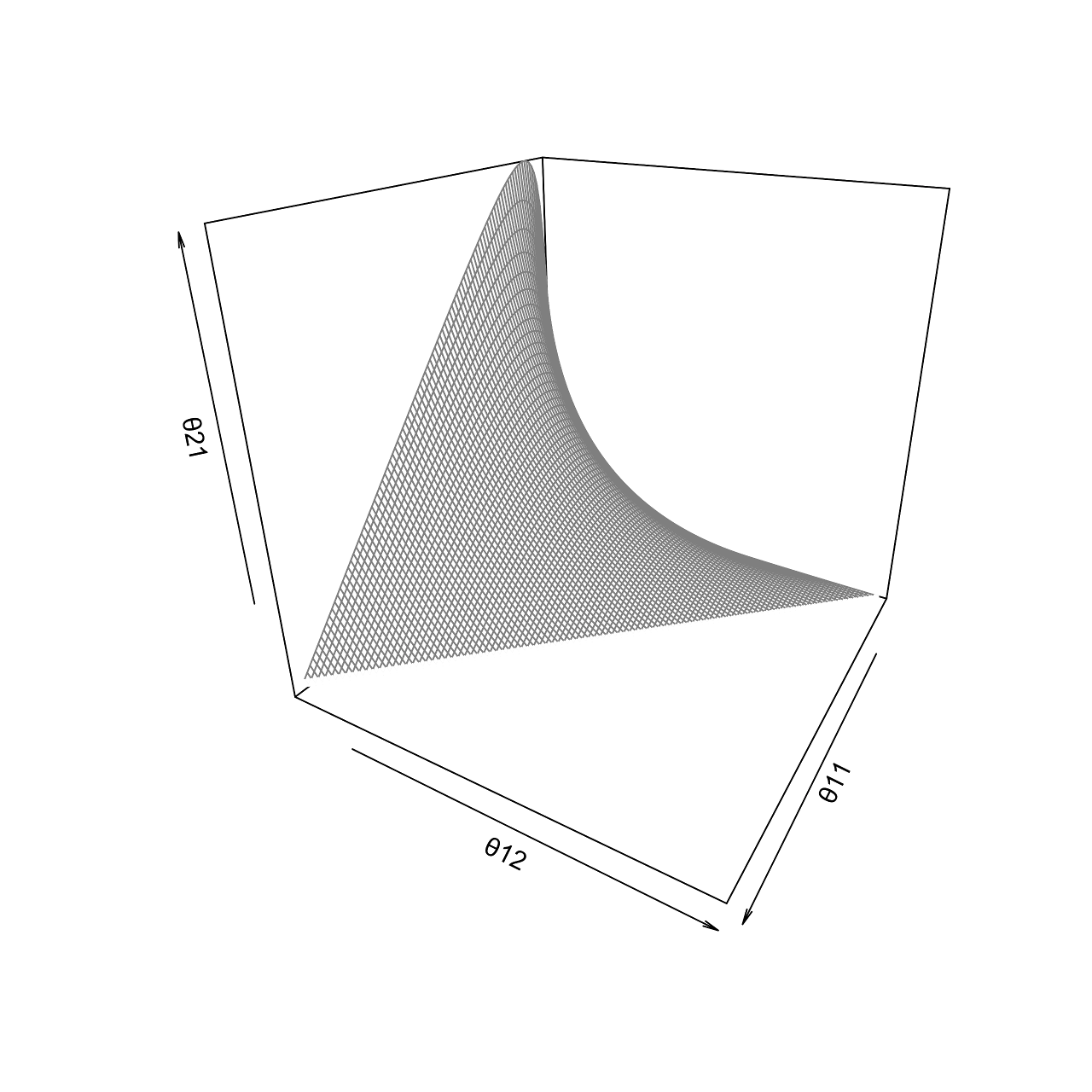}
\caption{Independence hypothesis (gray surface) for $2 \times 2$ tables and parametric space is the three-dimensional simplex (regular tetrahedron). \label{fig_2}}
\end{figure}

For the independence hypothesis, our interest is to test $\bm{H}: \theta_{ij} = \theta_{i\cdot} \times \theta_{\cdot j}$, $\forall i, j$. For the case of $2 \times 2$ table, the independence hypothesis is geometrically represented as Figure \ref{fig_2}. 

Considering ~that $n_{\cdot\cdot}$ is known, we ~assume that the outcomes ~of Table \ref{tab_vero2} follow a \mbox{$\textnormal{Multinomial}(n_{\cdot\cdot}, \bm{\theta})$} distribution, $\bm{\theta} = \{\theta_{11}, \ldots, \theta_{1(c-1)}, \dots, \theta_{\ell 1}, \ldots, \theta_{\ell(c-1)}\}$, and $\theta_{ic} = 1-\sum_{j=1}^{c-1} \theta_{ij}$. The likelihood function is
\[ L(\bm{\theta} \mid \bm{x}) = \left[ \left. n_{\cdot \cdot}! \middle/ \prod\limits_{i=1}^\ell \prod\limits_{j=1}^c x_{ij}! \right. \right] \prod\limits_{i=1}^\ell \prod\limits_{j=1}^c \theta_{ij}^{x_{ij}}. \]

The likelihood function under $\bm{H}$ is
\[ L(\bm{\theta} \mid \bm{x}, \bm{H}) = \left[ \left. n_{\cdot \cdot}! \middle/ \prod\limits_{i=1}^\ell \prod\limits_{j=1}^c x_{ij}! \right. \right] \prod\limits_{i=1}^\ell \theta_{i \cdot}^{n_{i\cdot}} \prod\limits_{j=1}^c \theta_{\cdot j}^{n_{\cdot j}}, \]
and the LRT $\lambda$ statistic is
\begin{equation}
\label{eq_lambdaind}
\lambda(\bm{x}) = \left. \prod\limits_{i=1}^\ell \left(\frac{n_{i \cdot}}{n_{\cdot \cdot}}\right)^{n_{i\cdot}} \prod\limits_{j=1}^c \left(\frac{n_{\cdot j}}{n_{\cdot\cdot}} \right)^{n_{\cdot j}} \middle/ {\prod\limits_{i=1}^\ell \prod\limits_{j=1}^c \left(\frac{x_{i j}}{n_{\cdot\cdot}} \right)^{x_{i j}}} \right. .
\end{equation}

$\bullet$ {\bf{P-value:}}\\
As shown in Subsection \ref{sec_homo22}, the P-value is obtained the same way but with a different $h(\bm{x})$. In this case, 
\[ h(\bm{x}) = \left. n_{\cdot \cdot}!(n_{\cdot \cdot}+\ell)! (n_{\cdot \cdot}+c)! \middle/ \left[ \prod\limits_{i=1}^\ell \prod\limits_{j=1}^c n_{ij}! \prod\limits_{i=1}^\ell n_{i \cdot}! \prod\limits_{j=1}^c n_{\cdot j}! \right] \right. . \]

~\\
$\bullet$ {\bf{FBST:}}\\
Assuming ~~a ~$\textnormal{Dirichlet}(1,1,\ldots,1)$ ~as ~~prior ~distribution for $\bm{\theta}$ ~and that ~~the ~outcomes ~of ~Table \ref{tab_vero2} follow a $\textnormal{Multinomial}(n,\theta_{11},\theta_{12},\ldots,\theta_{\ell 1},\ldots,\theta_{\ell c})$ distribution, then the posterior distribution is a $\textnormal{Dirichlet}(x_{11}+1,x_{12}+1,\ldots,x_{\ell 1}+1, \ldots, x_{\ell c}+1)$. The e-value is obtained from Definition \ref{def_e} and
\[ \sup_{\bm{\theta} \in \Theta_{\bm{H}}} \pi(\bm{\theta} \mid \bm{x})
= \frac{x_{11}!x_{12}!\cdots x_{\ell 1}! \cdots x_{\ell c}}{(n+\ell c -1)!} \prod\limits_{i=1}^\ell \left( \frac{n_{i \cdot}}{n}\right)^{n_{i \cdot}} \left( \frac{n_{ \cdot j}}{n}\right)^{n_{\cdot j}} . \]

~\\
$\bullet$ {\bf{Other indices:}}\\
We obtained the asymptotic LRT p-value and e-value, considering that $-2ln(\lambda(\bm{X}))$ follows a $\mathcal{X}^2$ distribution with $\ell+c-2$  and  $(\ell \times c)-1$ degrees of freedom. We also obtained the p-value for the Chi-Square independence test. 

\subsection{Hardy-Weinberg equilibrium}
\label{sec_hw}

An individual's genotype is formed by a combination of alleles. If there are two possible alleles for one characteristic (say $A$ and $a$), the possible genotypes are $AA$, $Aa$ or $aa$. Considering a few premises true \citep{Hartl2007}, the principle says that the allele probability in a population does not change from generation to generation. It is fundamental for Mendelian mating by allelic model. 
If the probabilities of alleles are $\theta$ and $1-\theta$, the expected genotype probabilities  are $(\theta^2, 2\theta(1-\theta), (1-\theta)^2)$ $0 \leq  \theta \leq 1$.

\begin{figure}
\centering
  \includegraphics[width=0.45\textwidth,keepaspectratio=true]{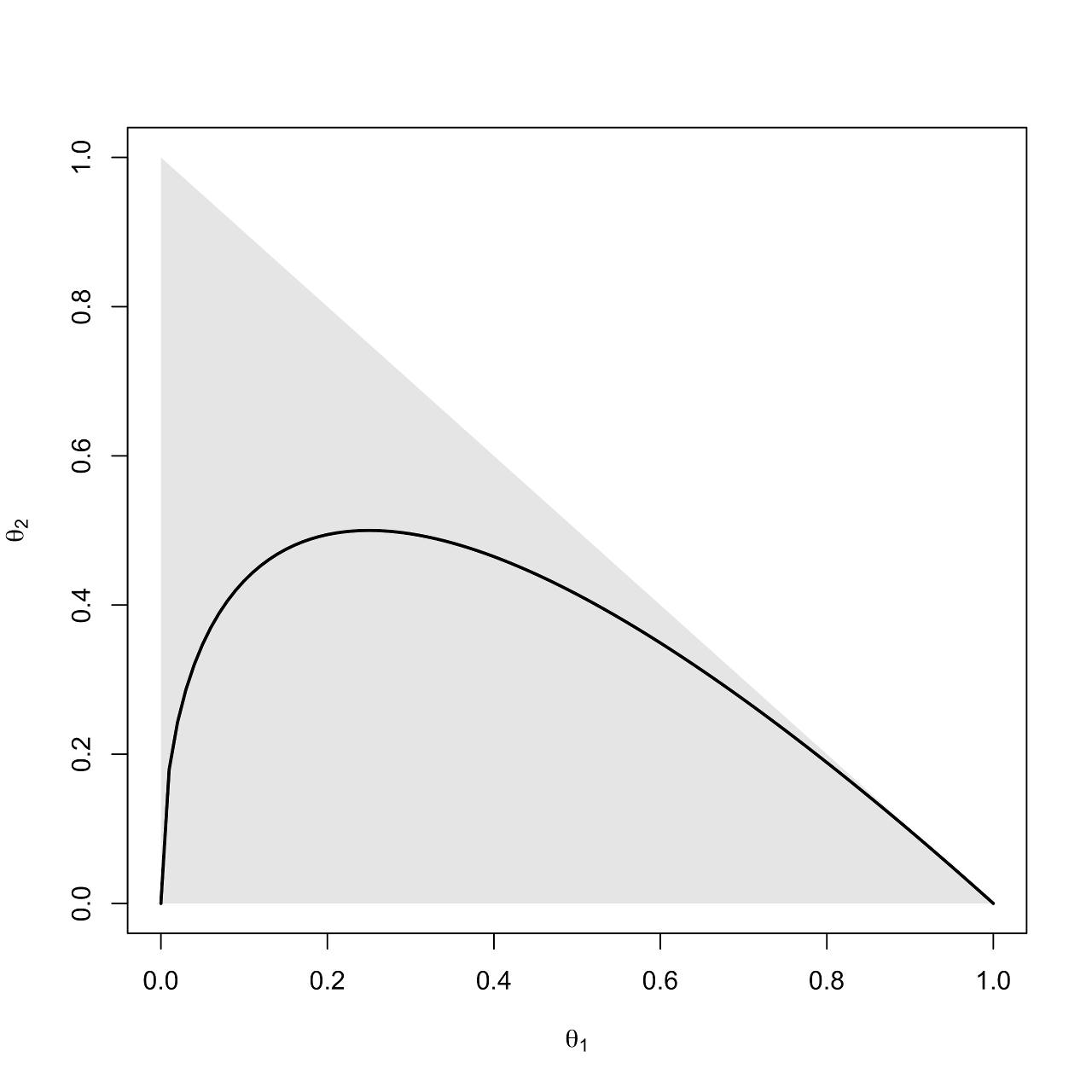}
\caption{Hardy-Weinberg equilibrium hypothesis (black line) and parametric space (gray shading). \label{fig_3}}
\end{figure}

Considering the Hardy-Weinberg equilibrium, the aim is to verify if a population follows these genotypes proportions. Therefore, the equilibrium hypothesis is
\[ \bm{H}: \left\{\begin{array}{rcl}
\theta_1 & = & \theta^2,\\
\theta_2 & = & 2\theta(1-\theta),\\
\theta_3 & = & (1-\theta)^2,
\end{array} \right. \]
in which $\theta_1, \theta_2, \theta_3$ are the proportions of AA, Aa, and aa, respectively. This hypothesis is geometrically represented in Figure \ref{fig_3}. 

Let $X$ be a random vector. Table \ref{tab_hw} represents the genotype frequencies for the population in question. Considering $n$ known, we assume that $X$ follows a $\textnormal{Trinomial}(n, \theta_1, \theta_2, \theta_3)$ distribution. The likelihood function for this model is
\[L(\bm{\theta} \mid \bm{x}) = \left[ \frac{n!}{x_1!x_2!x_3!} \right] \theta_{1}^{x_{1}}\theta_{2}^{x_{2}}\theta_{3}^{x_{3}}, \theta_{1} + \theta_{2} +\theta_{3} = 1, \]
and under the hypothesis $\bm{H}$, 
\begin{eqnarray*}
L(\bm{\theta} \mid \bm{x}, \bm{H}) & = & \left[ \frac{n!}{x_1!x_2!x_3!} \right] 2^{x_2}\theta^{2x_{1}+x_2} (1-\theta)^{2x_{3}+x_2},  0 \leq \theta \leq 1.
\end{eqnarray*}

\begin{table}[!ht]
\centering
\caption{\label{tab_hw} Genotype frequency.}
\begin{tabular}{c|ccc|c}
\hline
         &  $AA$       &  $Aa$       &    $aa$    & total \\
\hline
   $X$ &  $x_{1}$    &  $x_{2}$  &    $x_{3}$ & $n$  \\
\hline
\end{tabular}
\end{table}

The maximum likelihood estimator for $\theta$ under $\bm{H}$ is $\hat{\theta}= \frac{2x_1+x_2}{2(x_1+x_2+x_3)}$ and the LRT $\lambda$ statistic is
\begin{equation}
\label{eq_lambdahw}
\lambda(\bm{x}) = \frac{2^{x_2}\hat{\theta}^{2x_1+x_2}(1-\hat{\theta})^{2x_3+x_2}}{{(\frac{x_1}{n})}^{x_{1}}{(\frac{x_2}{n})}^{x_{2}}{(\frac{x_3}{n})}^{x_{3}}}.
\end{equation}

~\\
$\bullet$ {\bf{P-value:}}\\
Calculations follow as for the other indices and in this scenario
\[h(\bm{x}) = \frac{n!2^{x_2}(2x_1+x_2)!(2x_3+x_2)!}{x_1!x_2!x_3!(2x_1+2x_2+2x_3+1)!}.\]

~\\
$\bullet$ {\bf{FBST:}}\\
Assuming a $\textnormal{Dirichlet}(1,1,1)$ prior for $\bm{\theta}$ and that $X$ follows a \mbox{$\textnormal{Trinomial}(n,\theta_1,\theta_2,\theta_3)$} distribution, the posterior distribution is $\bm{\theta} \mid \bm{x} \sim \textnormal{Dirichlet}(x_{1}+1,x_{2}+1,x_{3}+1)$. In this setting,
\[ \sup_{\bm{\theta} \in \Theta_{\bm{H}}} \pi(\bm{\theta} \mid \bm{x})
= \frac{x_{1}!x_{2}!x_{3}!}{n!} 2^{2 x_2} \left(\frac{2x_1+x_2}{2n}\right)^{2x_1+x_2} \left(1 - \frac{2x_1+x_2}{2n}\right)^{x_2+2x_3}. \]

~\\
$\bullet$ {\bf{Other indices:}}\\
Both asymptotic LRT p-value and asymptotic e-value are obtained, the p-value considering that $-2\ln(\lambda(\bm{X}))$ follows a $\mathcal{X}^2$ distribution with $1$ degrees of freedom and the FBST considering that it follows a $\mathcal{X}^2$ distribution with $2$ degrees of freedom. The Chi-Square test was also performed in this scenario. 

\section{Relations between the indices}
\label{sec_compare}

As our objective is to compare the indices, we consider different scenarios for each hypothesis. For each scenario, we evaluate the significance indices of all test procedures presented in previous sections. Note that this is not a simulation study; for each sample size, we evaluate the indices for all possible contingency tables of a fixed dimension and size. For example, considering homogeneity hypothesis in a $2 \times 2$ table with marginals $(10,10)$, there are 121 possible tables or considering independence hypothesis in a $2 \times 3$ table with marginal $15$, there are 15504 possible tables. We evaluated the indices for all the tables that fit into each specification. For the e-value computation, non-informative priors for the parameters are considered (that is, $\pi(\bm{\theta}) \propto 1$). This way, no extra information is added besides the data, allowing fair comparisons between frequentist and Bayesian indices. 

In many practical situations, mainly in biological studies, asymptotic distributions are used to evaluate indices even for small samples. With that in mind, one of our interests is to understand how the use of asymptotic results for small sample size settings compares to the use of an exact index. Surprisingly, the values of exact and asymptotic indexes do not diverge considerably. 

For each scenario, plots are drawn to illustrate possible differences among the values of the indices. The indices studied are the P-value, asymptotic p-value for the LRT, asymptotic p-value for the chi-square test, e-value and asymptotic e-value. For the homogeneity hypothesis in $2 \times 2$ tables, Fisher's exact test was also obtained. We considered many different scenarios, however, since the aim is to understand the indices in small sample size, the scenarios presented here are in Table \ref{tab_compare}.

\begin{table}[ht]
\centering
\caption{\label{tab_compare} Considered scenarios.}
\begin{tabular}{cccc}
\hline
Setting &  Hypothesis       &  Table       &  Sample sizes    \\
\hline
1          &  Homogeneity    &  $2 \times 2$    &  $(30,30)$   \\
2          &  Homogeneity    &  $2 \times 2$    &  $(100,100)$   \\
3          &  Homogeneity    &  $2 \times 3$    &  $(30,30)$   \\
4          &  Homogeneity    &  $3 \times 3$    &  $(15,15,15)$   \\
5          &  Independence  &  $2 \times 2$    &  $30$   \\
6          &  Independence  &  $2 \times 3$    &  $30$   \\
7          &  Independence  &  $3 \times 3$    &  $15$   \\
8          &  Independence  &  $3 \times 3$    &  $25$   \\
9          &  Hardy-Weinberg equilibrium &  -    &  $30$   \\
10        &  Hardy-Weinberg equilibrium &  -    &  $100$   \\
\hline
\end{tabular}
\end{table}

\begin{figure}
\centering
  \includegraphics[width=\textwidth,keepaspectratio=true]{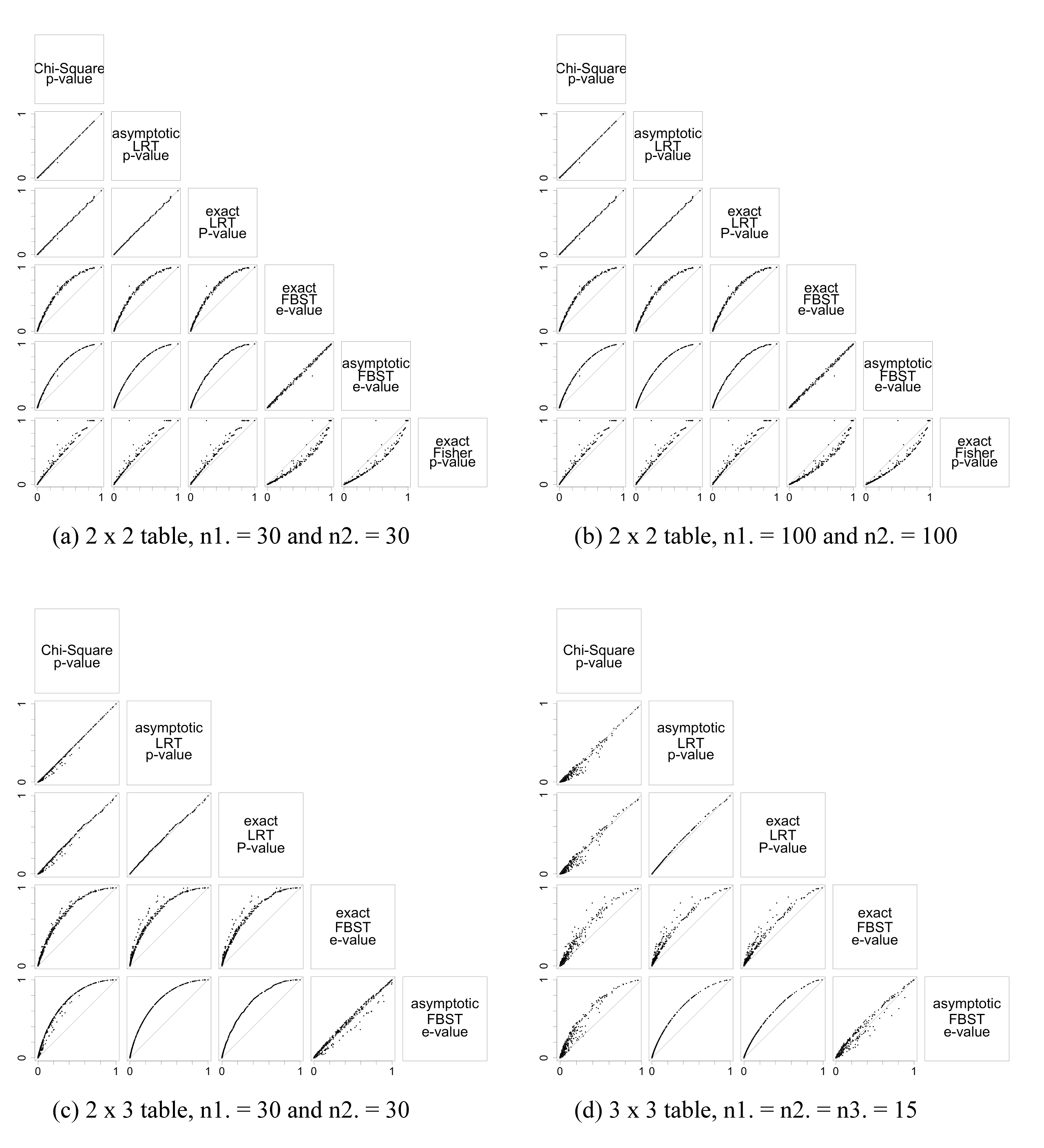}
\caption{Significance indices for homogeneity hypothesis considering different sample sizes and different table dimensions obtained for all possible tables. Each graph presents one index versus another, each dot representing a possible table, and if a dot is on top of the gray identity line, the two indices assume the same value for that table. \label{fig_4}}
\end{figure}

\begin{figure}
\centering
  \includegraphics[width=\textwidth,keepaspectratio=true]{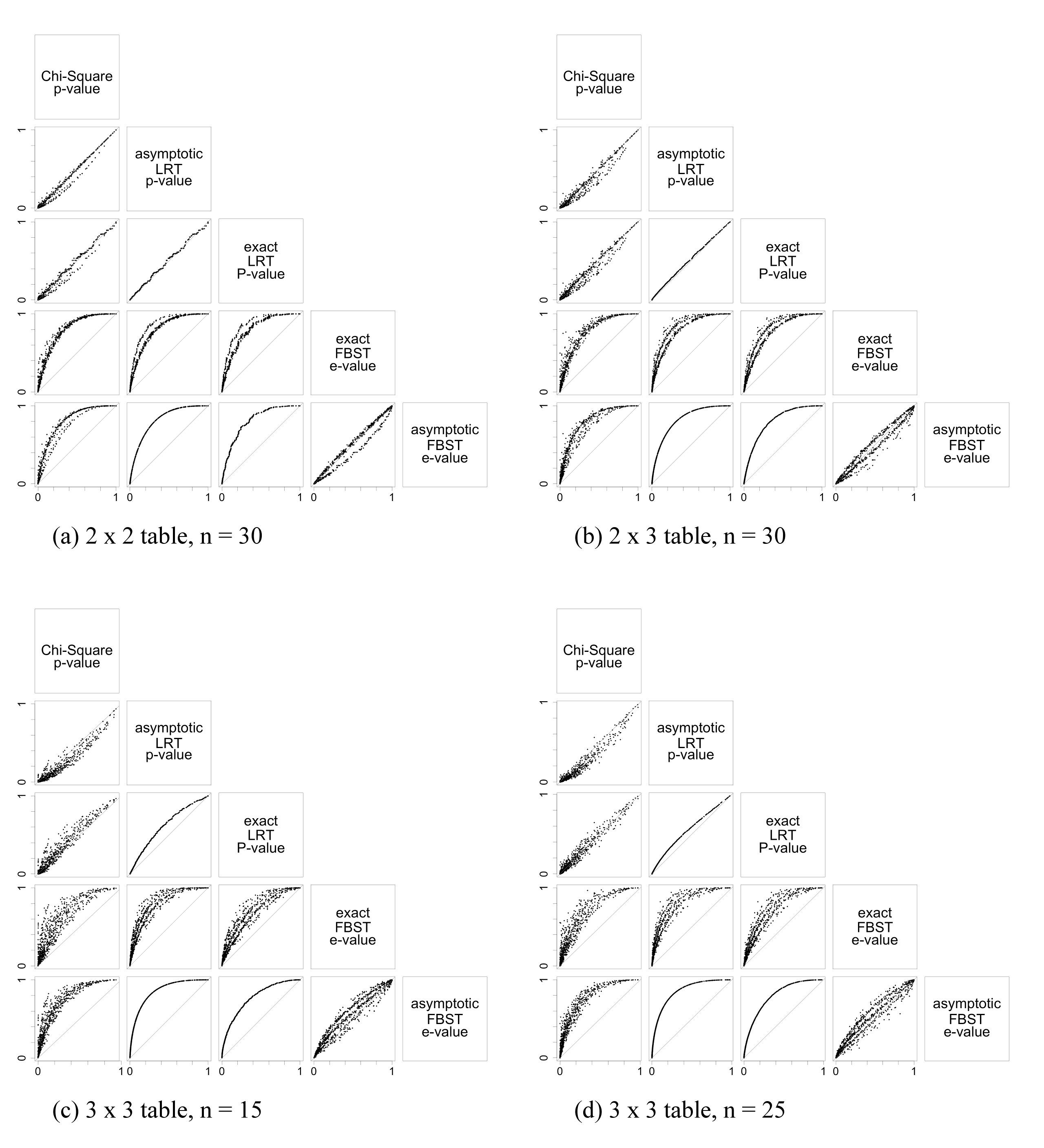}
\caption{Significance indices for independence hypothesis considering different sample sizes and different table dimensions obtained for all possible tables. Each graph presents one index versus another, each dot representing a possible table, and if a dot is on top of the gray identity line, the two indices assume the same value for that table. \label{fig_5}}
\end{figure}

\begin{figure}
\centering
  \includegraphics[width=\textwidth,keepaspectratio=true]{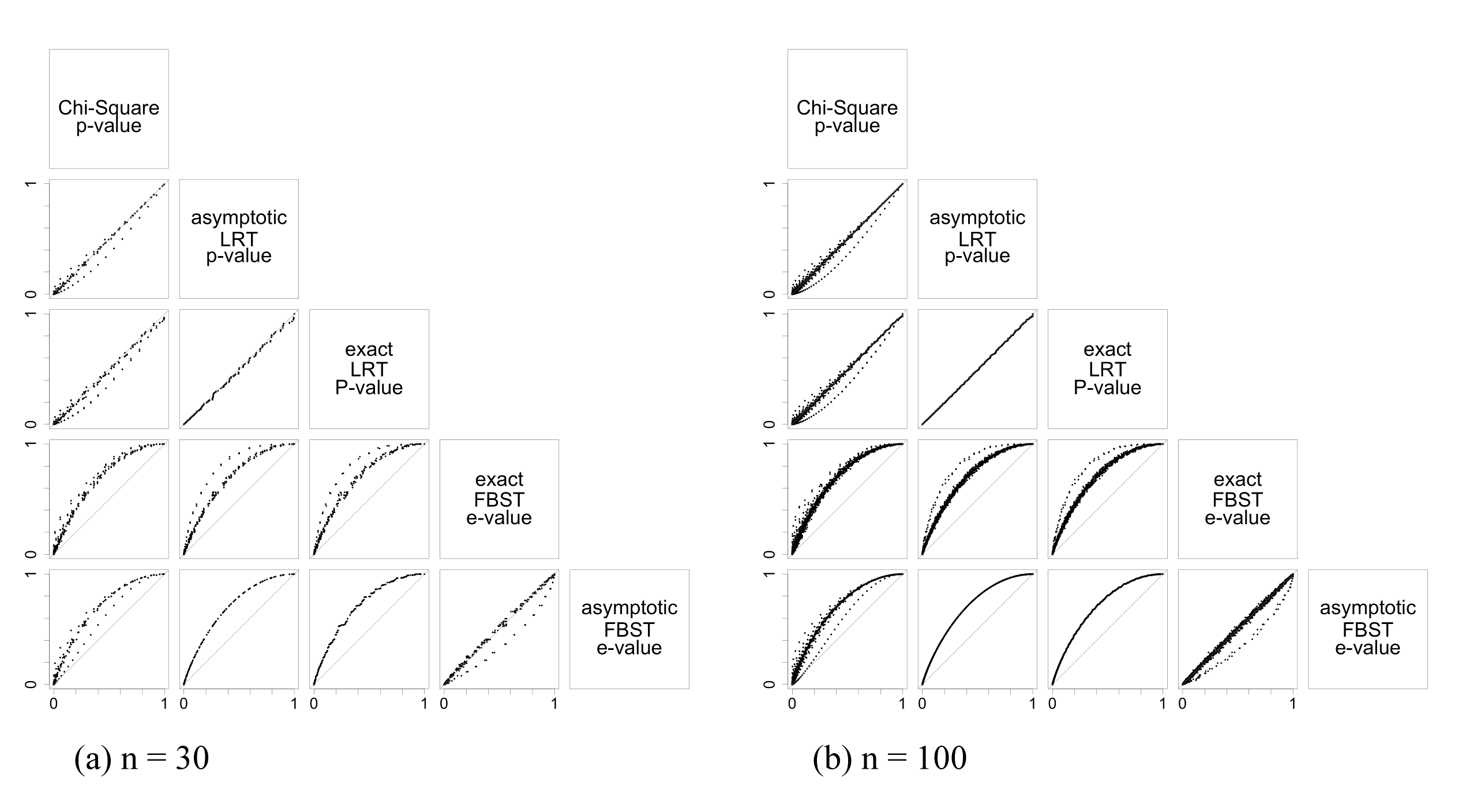}
\caption{Significance indices for Hardy-Weinberg equilibrium hypothesis considering different sample sizes and different table dimensions obtained for all possible tables. Each graph presents one index versus another, each dot representing a possible table, and if a dot is on top of the gray identity line, the two indices assume the same value for that table. \label{fig_6}}
\end{figure}

Figures \ref{fig_4}, \ref{fig_5} and \ref{fig_6} illustrate the results of the discussion above. For all hypotheses, exact and asymptotic e-values are very similar for both large and small sample sizes. In another direction, P-values and asymptotic p-values, both LRT and Chi-Square, are also very similar to each other. The difference found between e-values in comparison with both P-values and p-values happens as a result of the way these indices are developed. While e-values consider the full dimension of the parameter space (m degrees of freedom), P- and p-values consider the complementary dimension of the set corresponding to hypothesis $\bm{H}$ ($m - h$ degrees of freedom; $h$ is the dimension of the parameter sub-space defined by $\bm{H}$). This is expected from the asymptotic relationship between e-value and p-value from the LRT (Pereira et al. 2008, Diniz et al. 2012). Fisher's exact test was only calculated for the homogeneity hypothesis in $2 \times 2$ tables. It is the only index with a different behavior among the indices considered. This is not surprising, since it is a conditional exact test. Looking at the plots, its values do not form a continuous curve like the other indices' values do, and its points are quite far from all the other indices. 

The power function analyses of the frequentist tests for the homogeneity hypothesis in $2 \times 2$ table and Hardy-Weinberg equilibrium hypothesis is the object of the next section. 

\section{Power function}
\label{sec_power}

Power functions are a useful tool to compare hypothesis tests. For all $\theta \in \Theta$, the power function provides the probability of rejecting the hypothesis for a given $\theta$. In fact, we look for a test that does not reject the hypothesis for $\theta \in \Theta_{\bm{H}}$ and the further the $\theta$ value is from the hypothesis, the probability of rejection increases. 

The power functions presented are the ones that we are able to represent in $\mathbb{R}^3$, which are the power functions for the homogeneity hypothesis in $2 \times 2$ contingency tables and for the Hardy-Weinberg equilibrium hypothesis. 

We used p-values less than $0.05$ as a decision rule to reject the hypothesis. This choice is based on what is vastly used in most fields of science as a decision rule. In this case, $\textnormal{Power}(\theta_1,\theta_2) = P(\textnormal{reject $\bm{H}$}|(\theta_1,\theta_2))$ and
$\textnormal{Reject $\bm{H}$ if } \textnormal{index} \leq 0.05$.

We obtain the power function for all tests but the FBST. The FBST is a Bayesian significance test and in order to obtain a power function, one would need a decision rule. Since its construction differs from that of the p-values, we cannot use the same decision rule, and constructing a decision rule is not in the scope of this paper.

To the best of our knowledge, there is no analytic form for the power function of these tests, therefore we used a Monte Carlo procedure to evaluate it. We consider a grid for the unit square with $100 \times 100$ points on the axes $(\theta_1,\theta_2)$. For each point in the grid we generated 1000 tables. From these 1000 tables we evaluate the proportion of rejections, which is an approximation of the power function. 

\begin{figure}
\centering
  \includegraphics[width=\textwidth,keepaspectratio=true]{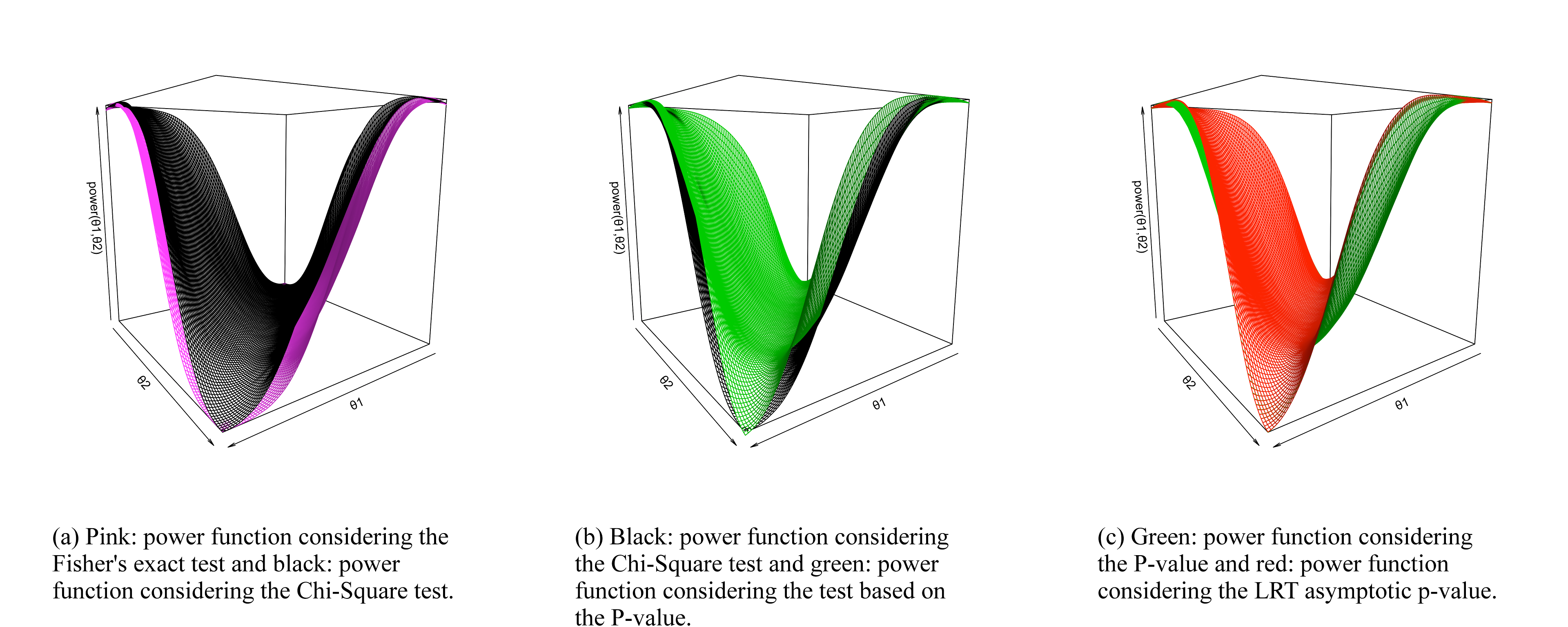}
\caption{Power function for homogeneity hypothesis in $2 \times 2$ contingency tables with $n_{1\cdot} = n_{2\cdot} = 10$. \label{fig_7}}
\end{figure}

\begin{figure}
\centering
  \includegraphics[width=0.9\textwidth,keepaspectratio=true]{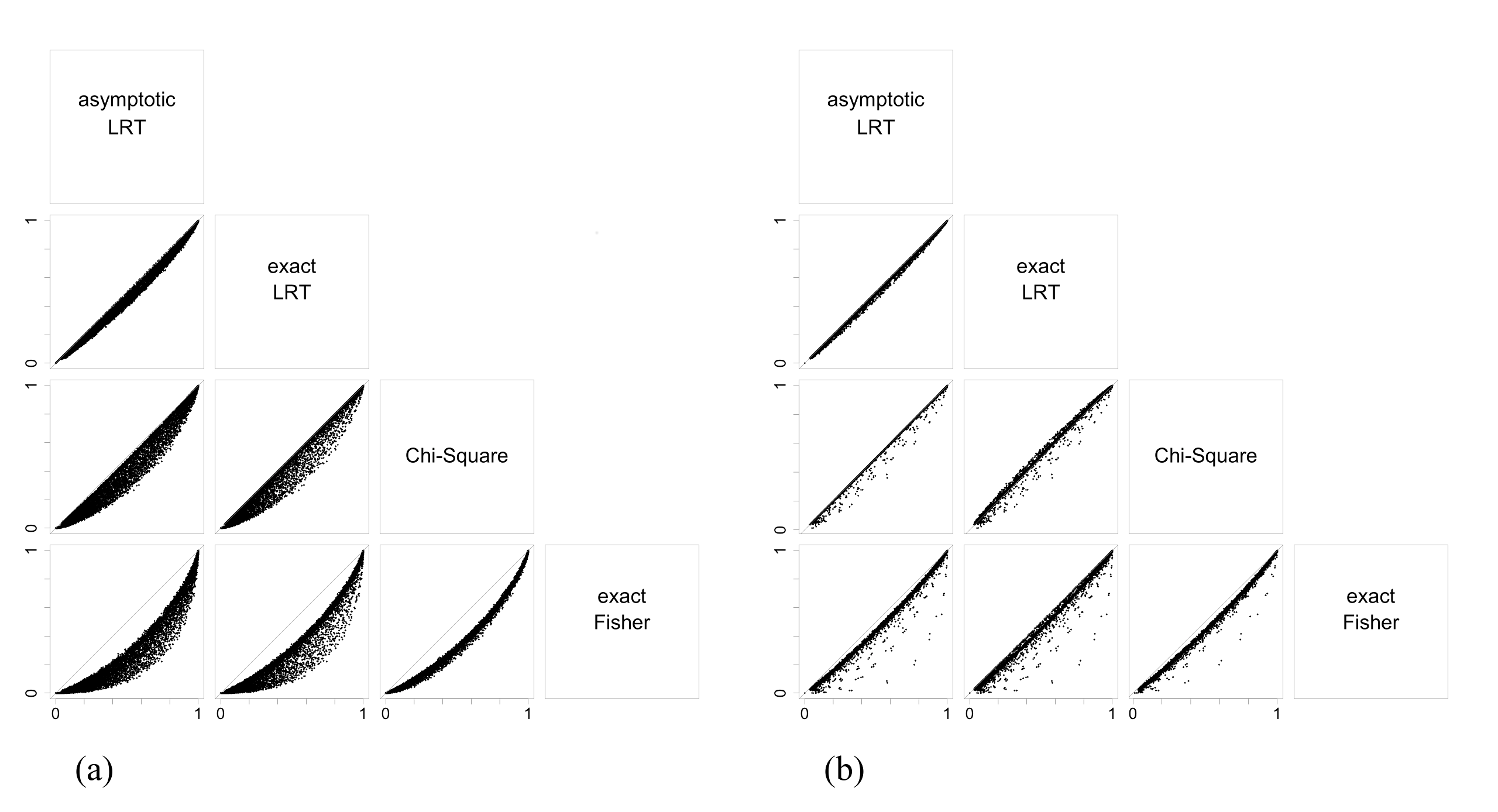}
\caption{Plots of power function values for the homogeneity test. Each graph presents one index versus another, each dot representing a point in the considered parametric space (in this case, $100 \times 100 = 10000$ points), and if a dot is on top of the gray identity line, the power functions assume the same value for that point in the parametric space. The scenario is $2 \times 2$ with marginals $n_{1\cdot} = n_{2\cdot} = 10$ in (a) and $n_{1\cdot} = n_{2\cdot} = 100$ in (b). \label{fig_8}}
\end{figure}

We plot pairs of power functions, in order to illustrate their shapes. For the homogeneity hypothesis in a table with marginals $(10,10)$, Figure \ref{fig_7} shows that Fisher's exact is less powerful than the Chi-square test, while the Chi-square is less powerful than the proposed P-value, which is less powerful than the asymptotic p-value for the LRT. To have a clear picture, we plot the power functions from different tests against each other. Figure \ref{fig_8}a consists of the power functions for tables with marginal equals to $(10,10)$. It shows that the use of the asymptotic p-value for the LRT results in a more powerful test than the other indices. When comparing the proposed P-value, it's more powerful than the Chi-square test and the Fisher's exact test. Between the Chi-square and the Fisher's test, the Chi-square test is more powerful. 

For tables with marginal equals to $(100,100)$, the graphs are more concentrated near the identity line (Figure \ref{fig_8}b), showing that all indices are more alike. The ordering still exists, but it is less severe. The only tests that show a different behavior than when considering marginals $(10,10)$ are the exact LRT (P-value) and the Chi-square test. They assume very similar values, all very close to the identity line, indicating that their power functions are very similar. It is interesting to point out that, as expected, the Chi-square test works better with larger samples.

\begin{figure}
\centering
  \includegraphics[width=0.8\textwidth,keepaspectratio=true]{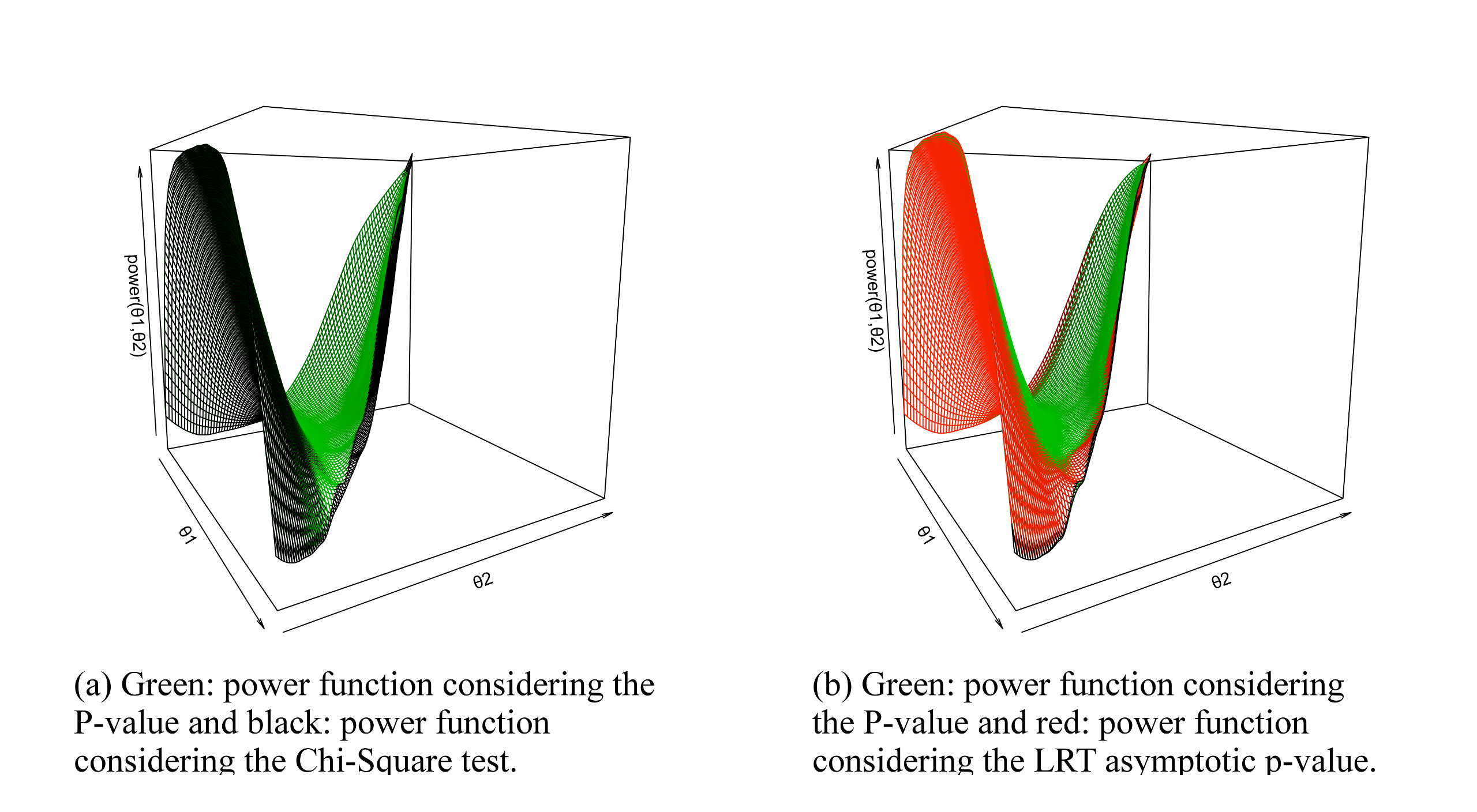}
\caption{\label{fig_9} Power function for Hardy-Weinberg equilibrium hypothesis with $n = 10$.}
\end{figure}

\begin{figure}
\centering
  \includegraphics[width=0.8\textwidth,keepaspectratio=true]{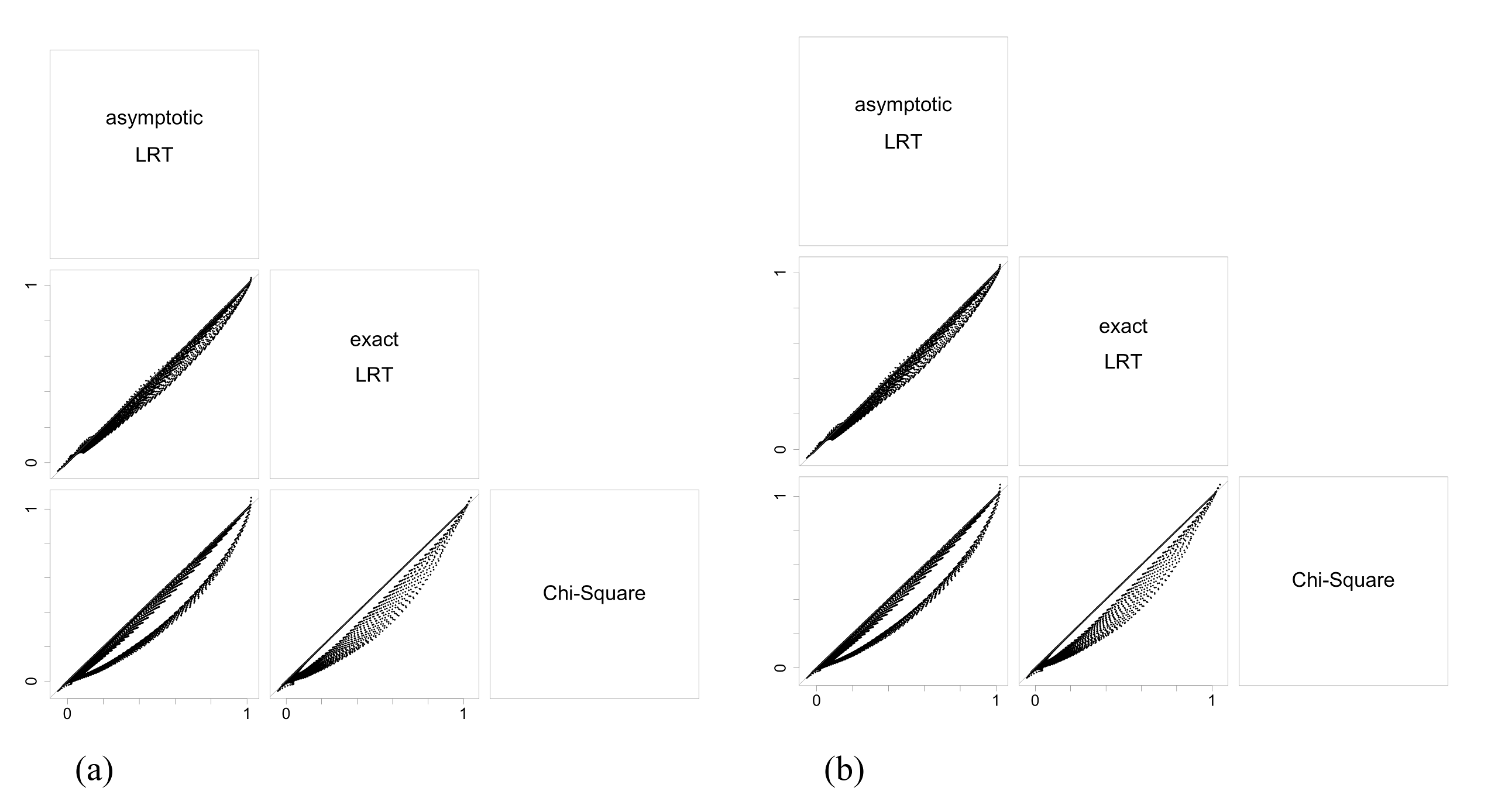}
\caption{Plots of power functions values for the Hardy-Weinberg equilibrium test. Each graph presents one index versus another, each dot representing a point in the considered parametric space (in this case, $100 \times 100 = 10000$ points), and if a dot is on top of the gray identity line, the power functions assume the same value for that point in the parametric space. The scenarios are marginals $n = 10$ (a) and $n =100$ (b). \label{fig_10}}
\end{figure}

For the Hardy-Weinberg hypothesis, the results are similar to the ones obtained for the homogeneity hypothesis and are shown in figures \ref{fig_9} and \ref{fig_10}. We call attention to the fact that, under hypothesis $\bm{H}$, the power function achieves the value of 0.05, as expected, since this is the significance level chosen to build the power functions.

\section{Discussion and Conclusion}
\label{sec_final}

After evaluating the indices for tables in different scenarios, we noticed that all of them had very similar behaviors, independently of the perspective (Bayesian or frequentist), sample size and table dimension. The exception is the p-value for Fisher's exact test for the homogeneity hypothesis in $2 \times 2$ tables, which shows distinct behavior. This seem to be a consequence of the drastic reduction of the sample space. Studying the power functions considering homogeneity hypothesis in $2 \times 2$ tables and Hardy-Weinberg equilibrium hypothesis, the LRT presented itself as the most powerful test when considering small sample sizes, while Fisher's exact test was the least powerful one for the homogeneity hypothesis and the Chi-Squares Test was the least powerful for the Hardy-Weinberg equilibrium hypothesis. By enlarging sample sizes, the power of these tests increases accordingly. 

Finally, we finish this paper listing our main conclusions: 
\begin{itemize}
\item The LTR asymptotic p-value seems to be a good frequentist alternative for small sample sizes.

\item Since there is an asymptotic relationship between the p-value for the LRT and the e-value (FBST), we consider that both indices are equivalent.

\item Taking into account available information besides the data, represented by informative priors, we consider the e-value a more appropriate index than a frequenstist one. 
\end{itemize}


\end{document}